\documentclass[12pt]{JHEP3}
\usepackage{amsmath,amssymb,bbm}
\usepackage{multirow,colordvi}
 \usepackage{epsf}
 \usepackage{epsfig}
 \usepackage{graphics}

\def\de{\partial}

\def\2{\frac12}
\def\4{\frac14}

\def\a{\alpha}
\def\b{\beta}
\def\g{\gamma}
\def\G{\Gamma}
\def\d{\delta}
\def\e{\epsilon}

\def\l{\lambda}
\def\L{\Lambda}
\def\m{\mu}
\def\n{\nu}

\def\r{\rho}

\def\S{\Sigma}

\def\de{\partial}

\def\be{\begin{equation}}
\def\ee{\end{equation}}
\def\bea{\begin{eqnarray}}
\def\eea{\end{eqnarray}}

\author{Eric A. Bergshoeff, Mees de Roo, Sven F. Kerstan\\Centre for
Theoretical Physics, University of
Groningen, Nijenborgh 4, 9747 AG Groningen, The Netherlands\\
\email{E.A.Bergshoeff, M.de.Roo, S.Kerstan@rug.nl}}
\author{Tom\'as Ort\'\i n\\Instituto de F\'\i sica Te\'orica
UAM/CSIC, Facultad de Ciencias C-XVI, C.U. Cantoblanco,
E-28049-Madrid, Spain\\ \email{Tomas.Ortin@cern.ch} }
\author{Fabio Riccioni
\\ DAMTP, Centre for Mathematical Sciences,
University of Cambridge,  Wilberforce Road, Cambridge CB3 0WA, UK;\\
Dipartimento di Fisica, Universit{\`a} di Roma ``Tor Vergata'',
I.N.F.N. - Sezione di Roma II ``Tor Vergata'', Via della
Ricerca Scientifica, 1 - 00133 Roma - ITALY\\
\email{F.Riccioni@damtp.cam.ac.uk}}

\abstract{We show that IIA supergravity can be extended with
two independent 10-form potentials. These give rise to a single BPS
IIA 9-brane. We investigate the bosonic gauge
algebra of both IIA and IIB
supergravity in the presence of 10-form potentials
and point out an intriguing relation with the symmetry
algebra $E_{11}$, which has been conjectured to be the underlying symmetry of
string theory/M-theory.}

\title{IIA Ten-forms and the Gauge Algebras of
Maximal Supergravity Theories}

\preprint{UG-06-03\\
IFT-UAM/CSIC-06-04\\
DAMTP-2005-111\\
ROM2F/06-04}

\begin{document}

\section{Introduction\label{Intro}}
Ten-dimensional type-IIB superstring theory is conjectured to
possess an $SL(2,\mathbb{Z})$ self-duality \cite{hulltownsend}. This
non-perturbative symmetry, which is a discrete subgroup of the
$SL(2,\mathbb{R})$ symmetry of the low-energy effective action \cite{IIB},
transforms the various BPS branes in the theory. While the D1-brane
and D5-brane solutions belong to a doublet, and the D3-brane to a
singlet, the  D7-brane solution of \cite{ggp} transforms non-linearly
with respect to $Sl(2,\mathbb{Z})$. Its charge matrix has vanishing
determinant~\cite{mo}, while half-supersymmetric
7-brane solutions in other conjugacy
classes~\cite{7branes} can be obtained as bound states of the
(anti-) D7-brane and its S-dual.
It turns out that the 7-branes transform as
a nonlinear doublet under $SL(2,\mathbb{Z})$.
Recently, the leading terms of a kappa-symmetric action for these
7-branes, involving their tensions, have been derived~\cite{9branes}.

Type-IIB string theory also possesses D9-branes, that identify the
open sector of the type-I theory, obtained from a projection of
IIB~\cite{augusto} known as {\it orientifold projection}. In the
closed sector, this projection corresponds to the insertion of
O9-planes, whose charge has to cancel that of the D9-branes.
Although the presence of D9-branes is not consistent if the overall
charge is not canceled, it is  possible to write a
kappa-symmetric effective action for these objects, whose
Wess-Zumino term contains a coupling to a RR 10-form. The gauge and
supersymmetry transformations for this form were derived
in~\cite{D9truncations}. From a careful analysis of the
supersymmetry algebra, it was shown in~\cite{BdRKR} that this
10-form belongs to a quadruplet of $SL(2,\mathbb{R})$.
Requiring the
leading terms in the corresponding effective action to be invariant
under 16 linear supersymmetries leads to a
constraint on the charges, so that only a {\it non-linear} doublet
of 9-branes remains. Furthermore, the theory
contains an additional {\it linear} doublet of 10-forms.
The 10-forms in this linear doublet
give rise to supersymmetric effective actions without the
need to impose constraints~\cite{9branes}.

One of the aims of this paper is to make for IIA supergravity the same
analysis that was done for IIB in \cite{BdRKR}.
In
\cite{domainwalls} a {\it democratic} formulation for IIA was given,
in which all the RR forms were considered together with their
magnetic duals. The resulting supersymmetry algebra has the
feature of describing both the `massless' IIA
supergravity \cite{IIA} and Romans' massive theory \cite{romans},
whose cosmological constant is treated as the dual of a 10-form
field strength, whose 9-form RR potential couples to D8-branes
\cite{pw,bdrgpt}. This 9-form does not carry propagating degrees of
freedom, and is therefore not dual to propagating supergravity
fields. A further dualization, now also including the dilaton and the NSNS
2-form, was performed in \cite{Juliaetal} (see also \cite{bandos}).
In~\cite{fabio} it was shown that the IIA supergravity theory can be
extended in order to include a 10-form. The
corresponding spacetime-filling brane has a tension scaling like
$g_S^{-2}$ in the string frame, and it is the T-dual \cite{fabio} of
a similar solitonic IIB 9-brane in the linear doublet. In order to
determine all the possible 10-forms in IIA supergravity, we
perform an analysis analogous to the one in \cite{BdRKR}.
In particular,  we construct
a completely democratic formulation, in which all the
fields, and not only the RR ones, are introduced together with their
magnetic duals. The outcome of this analysis will be that there are
two independent 10-forms in the IIA theory. We also analyze which of these
10-forms can give rise to a kappa-symmetric 9-brane action.

It turns out that the 10-forms implied by supersymmetry in both IIA
and IIB supergravity are  the ones that are predicted by
$E_{11}$~\cite{kleinschmidtwestschnakenburg,west10forms}, a
conjectured infinite-dimensional symmetry underlying string and
M-theory~\cite{west1,west3,west2}. {Related approaches with extended
symmetry algebras have been discussed in
\cite{damour,englert,nicolai}.} In this paper we wish to further
discuss the {intriguing relationship between the $E_{11}$ approach
and results derived from supersymmetry}. In particular, we will show
that, after a suitable (field-dependent) redefinition of the gauge
fields and the gauge parameters, the gauge transformations of all
the forms become {\it linear} in the gauge fields, while the
resulting bosonic gauge algebra is {\it
non-Abelian}~\cite{Juliaetal}. We perform this analysis for both IIA
and IIB supergravity.

The structure of the paper is as follows. In Section \ref{IIA} we write
IIA supergravity in a completely democratic formulation, and we show that
the IIA theory allows two independent 10-forms. In Section \ref{Brane}
we show that the two 10-forms can give rise to a single
kappa-symmetric IIA 9-brane.
Section \ref{bosonicalgebras} is devoted to an
analysis of the bosonic gauge algebras of both IIA and IIB supergravity.
In particular, we point out an intriguing relationship between
the commutation rules
for the gauge transformations and certain predictions from
$E_{11}$. In section \ref{M-theory}
we discuss $E_{11}$ and M-theory.
Finally, section \ref{Concs} contains the conclusions.

\section{IIA Supergravity and Ten-form Potentials\label{IIA}}

In this section we show that IIA supergravity allows two
independent 10-form potentials. We perform
the same analysis as was done in \cite{BdRKR} for the IIB case.
We use the notations and
conventions of \cite{domainwalls}, so that we will work in
string frame, with mostly plus signature. In this formulation, all
 RR fields and their magnetic duals are included, together with
the RR 9-form, whose field strength is dual to the cosmological
constant. This formulation describes both
the `massless' theory \cite{IIA} and Romans' theory \cite{romans}.
We will generalise this by including the
fields dual to the dilaton and the NSNS 2-form, that we call $B_{(8)}$
and $B_{(6)}$ respectively, and two 10-forms.

The propagating fields in the theory are the graviton $g_{\mu\nu}$,
the dilaton
$\phi$, the NSNS 2-form $B_{\m\n}$, the RR 1-form $C_\m$ and the RR
3-form $C_{\m\n\r}$, together with a Majorana non-chiral gravitino
$\psi_\m$ and a Majorana non-chiral dilatino $\l$ in the fermionic
sector. One then introduces the 7-form and 5-form duals of the RR
forms, together with a RR 9-form, whose 10-form field strength is
the dual of Romans' cosmological constant. Finally, supersymmetry
allows the introduction of at least one 10-form to this set of fields
\cite{fabio}. In the
rest of this paper, we will often denote  n-forms $F_{\m_1 \dots
\m_n}$ by $F_{(n)}$. Furthermore, antisymmetrization (with weight
one) of the indices is always understood. For instance, the expression $F_{(n)} G_{(m)}$ means
$F_{[\m_1 \dots \m_n} G_{\m_{n+1} \dots \m_{n+m}]}$. The same
notation will be used for gamma matrices, while the vielbein will be
denoted by $e^a$,  the gravitino by $\psi$ and the partial
derivative by $\de$. With $\Gamma_{11}$ we denote the
chirality matrix, defined as
    \be
  \G_{\m_1 \dots \m_{10}} = - \e_{\m_1 \dots \m_{10}} \G_{11} \quad
  .
  \ee
The supersymmetry transformations of all  fields to lowest
order in the fermions are \bea
\d e^a &=& \bar{\e}\, \G^a \psi\,, \\
\d B_{(2)} &=& 2 \bar{\e}\, \G_{11} \G_{(1)} \psi\,,\\
\d \phi &=& \tfrac{1}{2} \bar{\e}\, \l\,,\\
\d C_{(1)} &=& - e^{-\phi} \bar{\e}\, \G_{11} \psi  + \tfrac{1}{2}
  e^{-\phi} \bar{\e}\, \G_{11} \G_{(1)} \l\,, \\
\d C_{(3)} &=& -3 e^{-\phi} \bar{\e}\, \G_{(2)} \psi + \tfrac{1}{2}
  e^{-\phi} \bar{\e}\, \G_{(3)} \l + 3 C_{(1)} \d B_{(2)}\,,\\
\d C_{(5)} &=& -5 e^{-\phi} \bar{\e}\, \G_{11}\G_{(4)} \psi
  + \tfrac{1}{2} e^{-\phi} \bar{\e}\, \G_{11}\G_{(5)} \l + 10 C_{(3)}
\d B_{(2)}\,,\\
 \d C_{(7)} &=& -7 e^{-\phi} \bar{\e}\,\G_{(6)} \psi
  + \tfrac{1}{2} e^{-\phi} \bar{\e}\, \G_{(7)} \l + 21 C_{(5)}  \d B_{(2)}\,,\\
  \d C_{(9)} &=& -9 e^{-\phi} \bar{\e}\,\G_{11} \G_{(8)} \psi
  + \tfrac{1}{2} e^{-\phi} \bar{\e}\, \G_{11} \G_{(9)} \l + 36 C_{(7)}
\d B_{(2)}\,,\\
  \d {\cal D}_{(10)} &=& e^{-2\phi } ( -10 \bar{\e}\, \G_{(9)} \psi  +
\bar{\e}\,
  \G_{(10)} \l )\,,
  \label{first10form}
  \eea
for the bosons, while the fermions transform according
to\footnote{In the case of the fermions, we leave the index
structure explicit since contractions are involved.}
  \bea
  &&  \d \psi_\m = D_\m \e + \tfrac{1}{8} H_{\m\n\r} \G^{\n\r}
  \G_{11} \e + \tfrac{1}{8} e^{\phi} G^{(0)} \G_\m \e \nonumber \\
  & & \quad \quad + \tfrac{1}{16} e^{\phi} G_{\n\r} \G^{\n\r} \G_\m \G_{11}
  \e + \tfrac{1}{8 \cdot 4!} e^{\phi} G_{\m_1 \dots \m_4} \G^{\m_1
  \dots \m_4} \G_\m \e\,, \\
  & & \d \l = \de_\m \phi \G^\m \e - \tfrac{1}{12} H_{\m\n\r} \G_{11}
  \G^{\m\n\r} \e + \tfrac{5}{4} e^{\phi} G^{(0)} \e \nonumber \\
  & & \quad \quad + \tfrac{3}{8} e^{\phi} G_{\m\n} \G_{11} \G^{\m\n}
  \e + \tfrac{1}{4 \cdot 4!} e^{\phi} G_{\m_1 \dots \m_4} \G^{\m_1
  \dots \m_4} \e \quad .
  \eea
The bosonic gauge transformations are
  \bea
  \d B_{(2)} &=& 2 \de \S_{(1)}\,,\label{bgt1} \\
\d C_{(1)} &=& \de \L - G^{(0)} \S_{(1)}\,, \\
\d C_{(3)} &=& 3\de \L_{(2)} - H_{(3)} \L - 3G^{(0)} B_{(2)}\S_{(1)}\,, \\
\d C_{(5)} &=& 5\de \L_{(4)} - 10 H_{(3)} \L_{(2)} - 15G^{(0)} B_{(2)}^2
\S_{(1)}\,,\\
\d C_{(7)} &=& 7\de \L_{(6)} - 35 H_{(3)} \L_{(4)} - 105 G^{(0)} B^3_{(2)}
\S_{(1)}\,, \\
\d C_{(9)} &=& 9\de \L_{(9)} - 84 H_{(3)} \L_{(6)} - 945 G^{(0)} B^4_{(2)}
\S_{(1)}\,, \\
\d {\cal D}_{(10)} &=& 10 \de \S_{(9)}\quad ,\label{bgtl}
  \eea
and with respect to these, the field strengths
  \bea
  G_{(2)} &=& 2 \de C_{(1)} + G^{(0)} B_{(2)}\,,\label{fs1} \\
  H_{(3)} &=& 3 \de B_{(2)}\,, \\
  G_{(4)} &=& 4 \de C_{(3)} -4 H_{(3)}C_{(1)} + 3 G^{(0)} B^2_{(2)}\,, \\
  G_{(6)} &=& 6 \de C_{(5)} -20 H_{(3)}C_{(3)} + 15 G^{(0)}B^3_{(2)}\,, \\
  G_{(8)} &=& 8 \de C_{(7)} -56 H_{(3)}C_{(5)} + 105 G^{(0)}B^4_{(2)}\,, \\
  G_{(10)} &=& 10 \de C_{(9)} -120 H_{(3)}C_{(7)} + 945
  G^{(0)}B^5_{(2)} \label{fsl} \eea
are invariant. The various RR field strengths satisfy the duality
relations
  \be
  G^{(2n)}_{\m_1 \dots \m_{2n}} = (-1)^n \frac{1}{(10-2n)!}
  \e_{\m_1 \dots \m_{2n}}{}^{ \m_{2n+1}\ldots \m_{10}}
G^{(10-2n)}_{\m_{2n+1} \dots \m_{10}} \quad .
  \ee
In particular, the 10-form field strength is related to Romans'
cosmological constant $G^{(0)}$. For vanishing $G^{(0)}$, one
recovers the massless IIA supergravity theory.

Imposing the closure of the supersymmetry algebra, one can then
determine the supersymmetry transformation for the 6-form $B_{(6)}$,
whose field strength is related to $H_{(3)}$ by means of
  \be
  H_{\m_1 \dots \m_7 } = \tfrac{1}{6} e^{-2\phi} \e_{\m_1 \dots \m_7
  \m\n\r} H^{\m\n\r}\label{6formgct} \quad .
  \ee
The result is
  \bea
  \d B_{(6)}  &=&  6 e^{-2 \phi} \bar{\e}\, \G_{(5)} \psi - e^{-2\phi}
  \bar{\e}\, \G_{(6)} \l
  + 6 C_{(5)} \d C_{(1)} - 10 C_{(3)} \d C_{(3)} \nonumber\\
  &&-30 C_{(3)} B_{(2)} \d C_{(1)} + 30 C_{(3)} C_{(1)} \d B_{(2)} + 30
B_{(2)} C_{(1)} \d
  C_{(3)} \quad .
  \eea
The 7-form field strength reads
  \bea
  H_{(7)} &=& 7 \de B_{(6)} + G^{(0)} [ - \, C_{(7)}
  +\tfrac{105}{2} \, C_{(3)}B^2_{(2)} -\tfrac{105}{2} \, C_{(1)} B^3_{(2)} ]
\nonumber\\
  & & + G_{(2)} [ 21 \, C_{(5)} -105 \, C_{(3)} B_{(2)} ]
  + G_{(4)} [ -\tfrac{35}{2} \, C_{(3)} +\tfrac{105}{2} \, C_{(1)} B_{(2)} ]
  \quad ,
  \eea
and gauge invariance implies that $B_{(6)}$ transforms according to
  \bea
  \d B_{(6)}  &=& 6 \partial \Sigma_{(5)}
  + G^{(0)} [\L_{(6)} - \tfrac{45}{2} \L_{(2)} B^2_{(2)} +
  \tfrac{15}{2} \L B^3_{(2)}
  + 30 C_{(3)} \Sigma_{(1)} B_{(2)} \nonumber\\
  &&- 45 C_{(1)} \Sigma_{(1)} B^2_{(2)}]+ G_{(2)} [ -15 \L_{(4)}
  + 45 \L_{(2)} B_{(2)} -30 C_{(3)} \Sigma_{(1)}]\nonumber\\
  &&+ G_{(4)} [ \tfrac{15}{2}  \L_{(2)} - \tfrac{15}{2} \L B_{(2)}  +15
C_{(1)} \Sigma_{(1)}]\quad
  .
  \eea
Observe that if $G^{(0)}$ is non-zero, {\it i.e.} in the massive
theory, one can use $\L_{(6)}$ to gauge away $B_{(6)}$. This is
consistent with the fact that the 2-form becomes massive, because in
10 dimensions the dual of a massive 2-form is a massive 7-form, of
which $B_{(6)}$ describes the longitudinal components \cite{massive}.

Following the same strategy, we now determine the gauge and
supersymmetry transformations for the 8-form $B_{(8)}$ dual to the
dilaton. The duality relation is
  \be
  H_{\m_1 \dots \m_9} = e^{-2\phi} \e_{\m_1 \dots \m_9 \r} \de^\r
  \phi \quad ,
  \ee
while the  supersymmetry transformation turns out to be
  \bea
  \d B_{(8)} &=& \tfrac{1}{2} e^{-2\phi} \bar{\e}\, \G_{(8)} \G_{11} \l-6
C_{(7)} \d C_{(1)}
  +14 B_{(2)} \d B_{(6)} + 14 C_{(5)} \d C_{(3)} \nonumber\\
  &&- 210 B^2_{(2)} C_{(1)} \d C_{(3)}
  + 210 B^2_{(2)} C_{(3)} \d C_{(1)} - 42 C_{(5)} C_{(1)} \d B_{(2)}
  \quad .
  \eea
The 9-form field strength we find is
  \bea
  H_{(9)} & =& 9 \de B_{(8)} + G^{(0)} [\tfrac{5}{4} C_{(9)} -18
  C_{(7)} B_{(2)}
  + 315 C_{(3)} B^3_{(2)} - \tfrac{945}{4} C_{(1)} B^4_{(2)} ]\nonumber\\
  & & + G_{(2)} [ -27 C_{(7)} + 378 C_{(5)} B_{(2)} - 945 C_{(3)} B^2_{(2)}]\nonumber\\
  & & + G_{(4)} [\tfrac{63}{2} C_{(5)} - 315 C_{(3)} B_{(2)}
  + \tfrac{945}{2} C_{(1)} B^2_{(2)} ] -18 H_{(7)} B_{(2)}
  \quad ,
  \eea
and the 8-form gauge transformation is
  \bea
  \d B_{(8)} & =& 8 \de \S_{(7)}
  + G^{(0)} [ -\tfrac{5}{4} \L_{(8)} +14 \L_{(6)} B_{(2)}
  - 105  \L_{(2)} B^3_{(2)} + \tfrac{105}{4} \L B^4_{(2)}\nonumber \\
  & & -4 C_{(7)} \S_{(1)} + 210 C_{(3)} \S_{(1)} B^2_{(2)}
  - 210 C_{(1)} \S_{(1)} B^3_{(2)} ]\nonumber\\
  & & + G_{(2)} [ 21 \L_{(6)} - 210 \L_{(4)} B_{(2)} + 315 \L_{(2)} B^2_{(2)}
   + 84 C_{(5)} \S_{(1)} \nonumber \\
  & & - 420 C_{(3)} \S_{(1)} B_{(2)} ] + G_{(4)} [-\tfrac{ 35}{2} \L_{(4)} + 105 \L_{(2)} B_{(2)} - \tfrac{105}{2} \L B^2_{(2)}
   \nonumber \\
  &&-70 C_{(3)} \S_{(1)} + 210 C_{(1)} \S_{(1)} B_{(2)} ] -4 H_{(7)} \S_{(1)}
  \quad .
  \eea
As for the 6-form, in the massive theory, in which $G^{(0)}$
is non-vanishing, this 8-form can be gauged away by means of
$\L_{(8)}$. In this case this is related to the fact that in ten dimensions
the dual of a massive scalar is a massive 9-form potential.

Finally, we consider the inclusion of 10-forms.
Since these objects are not related
by duality to lower-rank fields, we can only use the closure of the
supersymmetry algebra to determine their gauge and supersymmetry
transformations. The final result is that, besides the 10-form
in eq. (\ref{first10form}), another 10-form $D_{(10)}$ can be
included in the algebra. Its supersymmetry transformation reads
  \bea
\d D_{(10)} &=& \tfrac{1}{2}e^{-2\phi} \bar{\e}\, \G_{(10)} \l
-\tfrac{15}{2} C_{(9)} \d C_{{(1)}}
    -45 B_{(2)} \d B_{(8)}+ 315 B^2_{(2)} \d B_{(6)} \nonumber \\
  && + \tfrac{63}{2} C_{(5)} \d C_{(5)} -315 C_{(3)} B_{(2)} \d C_{(5)} +
315 C_{(5)} B_{(2)} \d C_{(3)}\nonumber \\
  &&- 315 C_{(5)} C_{(3)} \d B_{(2)}+ \tfrac{945}{2} C_{(1)} B^2_{(2)} \d
C_{(5)}
  - \tfrac{945}{2} C_{(5)} B^2_{(2)} \d C_{(1)} \nonumber \\
  &&- 945 C_{(5)} C_{(1)} B_{(2)} \d B_{(2)}
   - 4725 C_{(1)} C_{(3)} B^2_{(2)} \d B_{(2)} \nonumber \\
  && -4725 C_{(1)} B^3_{(2)} \d C_{(3)}+ 4725 C_{(1)} B^4_{(2)} \d C_{(1)}
\label{delD10}
  \quad ,
  \eea
while the gauge transformation is
  \bea
\d D_{(10)} &=& 10 \de \S'_{(9)} + G^{(0)} [-2 \L_{(10)} +
\tfrac{225}{4} B_{(2)} \L_{(8)} - 315 B^2_{(2)} \L_{(6)}
  - \tfrac{1575}{4} B^3_{(2)} \L_{(4)}\nonumber \\
  &&+ \tfrac{4725}{2} B^4_{(2)} \L_{(2)} - \tfrac{945}{2} B^5_{(2)} \L
  - \tfrac{25}{2} C_{(9)} \S_{(1)} + 180 B_{(2)} C_{(7)} \S_{(1)}\nonumber\\
  &&+ \tfrac{945}{2} B^2_{(2)} C_{(5)} \S_{(1)} - 6300 B^3_{(2)} C_{(3)}
\S_{(1)}
  + 4725 B^4_{(2)} C_{(1)} \S_{(1)} ] \nonumber \\
&& + G_{(2)} [ \tfrac{135}{4} \L_{(8)} - 945 B_{(2)}
\L_{(6)}\nonumber
  + \tfrac{23625}{4} B^2_{(2)} \L_{(4)}
  -\tfrac{14175}{2} B^3_{(2)} \L_{(2)}  \nonumber \\
 && + 270 C_{(7)} \S_{(1)}
  -4725 B_{(2)} C_{(5)} \S_{(1)} + 14175 B^2_{(2)} C_{(3)} \S_{(1)}] \nonumber
\\
  && + G_{(4)} [\tfrac{1575}{4}  B_{(2)} \L_{(4)} -\tfrac{4725}{2}
  B_{(2)}^2 \L_{(2)} + \tfrac{4725}{4} B^3_{(2)} \L  \nonumber \\
 && - \tfrac{315}{2} C_{(5)} \S_{(1)} + 3150 B_{(2)} C_{(3)} \S_{(1)}
 - \tfrac{14175}{2} B^2_{(2)} C_{(1)} \S_{(1)} ] \nonumber \\
  && + G_{(6)} [ -\tfrac{105}{4} \L_{(4)}
  + \tfrac{315}{2} B_{(2)} \L_{(2)} -\tfrac{315}{4} B^2_{(2)} \L
  -105 C_{(3)} \S_{(1)} \nonumber \\
  &&+ 315 B_{(2)} C_{(1)} \S_{(1)} ] + 180 H_{(7)} B_{(2)} \S_{(1)} +
10 H_{(9)} \S_{(1)}
  \quad .\label{gauge10formfinal}
  \eea
The gauge parameter $\L_{(10)}$
(see the second term in the first line\footnote{
Observe that due to the presence of this
term, the 10-form $D_{(10)}$, like the 6-form and the 8-form, can be
 gauged
away in Romans' theory.})
plays a crucial role in closing the
algebra, and can be interpreted as the gauge parameter of an
11-form. If we allowed the dimension of spacetime to change
from $d=10$ to $d>10$ the 10-form $D_{(10)}$ generically
would describe
propagating degrees of freedom which would convert the 11-form into
a {\it massive} 11-form analogous to the massive 7-form and 9-form
we obtained above.
Since an 11-form is trivial in ten-dimensions, we are only left with the
10-form potential $D_{(10)}$. A similar phenomenon occurs in
the IIB case \cite{BdRKR}: the
field-strength of the (quadruplet of) 10-forms, considered formally
in $d>10$ dimensions,
contains non-trivial information about the gauge
transformations of potentials with rank higher than ten.
These observations hint at an underlying algebraic
structure which might be independent of the dimensionality of
space-time. We will discuss this structure in section
\ref{bosonicalgebras}.

A natural question to ask is whether the supersymmetry algebra allows
for the inclusion of additional 10-forms. The only  freedom we
have in the transformations to $\psi_\mu$ and $\lambda$ is to
change the dilaton factor, or to include an additional $\Gamma_{11}$
in the transformation rule. The last possibility 
leads to 10-forms for which the dilaton factor is not restricted 
by the supersymmetry algebra: these are all proportional to the 
ten-dimensional volume form and therefore not independent.
The possibility of changing the power of $e^{-\phi}$ without
including  $\Gamma_{11}$ in the transformation rules is ruled 
out by checking closure of the algebra. This analysis shows that
there are indeed only two independent 10-forms in the IIA supergravity
multiplet.

\section{Nine-branes of IIA\label{Brane}}

In \cite{9branes} we discussed the relation between the
1/2 BPS $p$-branes and $p+1$-form potentials
in IIB supergravity. In particular, we obtained the tensions
as well as the operator which projects onto the unbroken
linear supersymmetry. In this section we will do a similar
analysis for IIA supergravity, using the results of Section \ref{IIA}.

As an example we can work out the case of a 10-brane. We start
from the action (see \cite{9branes})
\begin{equation}\label{braneL}
{\cal{L}}_{\rm brane}\, =\, \tau_{\rm brane}\, \sqrt{-g} +
 x \epsilon^{\mu_1\cdots \mu_{10}}\,
{\cal D}_{\mu_1\cdots \mu_{10}}\,.
\end{equation}
Here we have assumed the existence of a (gauge-fixed)
kappa-symetric action, in static gauge.
Since $\tau_{\rm brane}$ will depend on the dilaton, the background
fields present in the Nambu-Goto term are metric and dilaton, the
Wess-Zumino term depends on the background potential. World-volume
fields play no role in this analysis. The action should exhibit 16
linearly realized supersymmetries. Therefore,
if we perform an $N=2$ supersymmetry transformation of the
background fields in (\ref{braneL}), we should find that half of the
supersymmetry parameters are projected out. It is sufficient to
consider the transformation from the bosonic fields to the gravitino
and dilatino. If $\tau_{\rm brane}$ is chosen correctly, the variation
to the gravitino will give a projection operator if the relative constant
between Nambu-Goto and Wess-Zumino terms is appropriately chosen. The
variation to the dilatino is the consistency check of this procedure.

In the present case the complete supersymmetry
variation of ${\cal D}_{(10)}$ is (\ref{first10form}):
\be
  \delta {\cal D}_{\m_1\ldots\m_{10}} = e^{-2\phi}\left(
  -10\bar\e\,\g_{[\m_1\ldots\m_{9}}\psi_{\m_{10}}
   + \bar\e\,\g_{[\m_1\ldots\m_{10}}\lambda\right)\,.
\label{delB10}
\ee
This determines the tension to be $e^{-2\phi}$.
The variation to the gravitino then fixes $x=1/10!$, while we find
for the projection operator $P=\tfrac{1}{2}(\mathbbm{1}+\G_{11})$. Using the same
value for $x$ the variation of ${\cal D}_{(10)}$ and dilaton to $\lambda$
produces the same projection operator.

The case of the 10-forms is particularly interesting because there is
a second 10-form, $D_{(10)}$, whose supersymmetry variation
is (\ref{delD10})
\be
 \delta D_{\m_1\ldots\m_{10}} = \tfrac{1}{2}e^{-2\phi}
   \left(\bar\e\,\g_{\m_1\ldots\m_{10}}\lambda
  + {\rm gauge-field\ dependent\ terms}\right)\,.
\ee
$D_{(10)}$ by itself cannot couple supersymmetrically to
a 9-brane, because there is no gravitino
contribution to match the variation of $\sqrt{-g}$ in (\ref{braneL}).
The result
is therefore that it is precisely ${\cal D}$, the
only combination which does not transform to gauge-field dependent
terms, that might correspond to a kappa-symmetric 9-brane.

In the IIB case we have a similar 10-form potential, which supersymmetrically
couples to a solitonic $(1/g_S)^2$ brane, and also does not transform to
gauge-field dependent terms \cite{9branes}. The absence of gauge fields
in the supersymmetry
transformation implies that these potentials have trivial bosonic gauge
transformations. This implies in turn that the Wess-Zumino term in
(\ref{braneL}) is gauge-invariant as it stands.

For completeness and further reference we present in Table \ref{Table1}
a list of all the BPS branes, their tension, potential and projection operator.
Note that also the NSNS-form $B_{(8)}$ is absent from the table, the reason
being that like $D_{(10)}$ it does not transform linearly to the gravitino.

\begin{table}[ht]
\begin{center}
\hspace{-1cm}
\begin{tabular}{|c||c|c|c|}
\hline \rule[-1mm]{0mm}{6mm}
potential & brane & tension &projection operator\\
\hline \rule[-1mm]{0mm}{6mm}
$ C_{(1)}$ & D0 & $e^{-\phi}$
&$\frac{1}{2}(\mathbbm{1}+\g_0)$\\
\hline \rule[-1mm]{0mm}{6mm}
$B_{(2)}$ & F1 & 1
&$\tfrac{1}{2}(\mathbbm{1} + \g_{01}\G_{11})$\\
\hline \rule[-1mm]{0mm}{6mm}
$C_{(3)}$ & D2  &   $e^{-\phi}$
& $\tfrac{1}{2}(\mathbbm{1} + \g_{012})$\\
\hline \rule[-1mm]{0mm}{6mm}
$C_{(5)}$ & D4  &   $e^{-\phi}$
&$\tfrac{1}{2}(\mathbbm{1} + \g_{01\ldots 4}\G_{11})$\\
\hline \rule[-1mm]{0mm}{6mm}
$B_{(6)}$ & NS5 &$e^{-2\phi}$
&$\frac{1}{2}(\mathbbm{1}+ \g_{01\ldots 5})$\\
\hline \rule[-1mm]{0mm}{6mm}
$C_{(7)}$ & D6  &   $e^{-\phi}$
&$\tfrac{1}{2}(\mathbbm{1}+\g_{01\ldots 6})$\\
\hline \rule[-1mm]{0mm}{6mm}
$C_{(9)}$ & D8  &   $e^{-\phi}$
&$\tfrac{1}{2}(\mathbbm{1} + \g_{01\ldots 8}\G_{11})$\\
\hline \rule[-1mm]{0mm}{6mm}
${\cal D}_{(10)}$ & NS9 &$e^{-2\phi}$
&$\frac{1}{2}(\mathbbm{1}+ \G_{11})$\\
\hline
\end{tabular}
\caption{Potentials, branes, tensions and projection operators for all
IIA supersymmetric branes.\label{Table1}}
\end{center}
\end{table}

\section{The Bosonic Gauge Algebras}\label{bosonicalgebras}

In this section we will analyse the algebra of bosonic gauge
transformations which is contained in the supersymmetry algebra. We
will first do this for IIB supergravity, then for IIA supergravity.
\\
Our analysis will reveal a surprising structure and a relation to
results from the $E_{11}$ approach
\cite{west1,west3,west2,kleinschmidtwestschnakenburg}. It may also
be seen as an extension and derivation from supersymmetry of the
results of \cite{Juliaetal}.

\subsection{The IIB Algebra}
Our starting point is the set of bosonic gauge transformations of IIB
supergravity in Einstein frame we obtained in \cite{BdRKR} where we used
a mostly minus signature:
\begin{eqnarray}\label{abelian}
\d A^\a_{(2)} &=& 2 \de \L^\a_{(1)}\quad\nonumber ,\\
\d A_{(4)} &=& 4 \de \L_{(3)} - \tfrac{i}{4} \e_{\g\d} \L^\g_{(1)}F^\d_{(3)}
\quad\nonumber ,\\
\d A^\a_{(6)} &=& 6 \de \L^\a_{(5)} - 8 \L^\a_{(1)} F_{(5)} - \tfrac{160}{3}
F^\a_{(3)}\L_{(3)}\quad\nonumber ,\\
\d A^{\a\b}_{(8)} &=& 8 \de \L^{(\a\b)}_{(7)} +\tfrac{1}{2}
F^{(\a}_{(7)} \L^{\b)}_{(1)}
  - \tfrac{21}{2} F^{(\a}_{(3)} \L^{\b)}_{(5)}\quad\nonumber ,\\
\d A^{\a}_{(10)} &=& 10 \de \L^{\a}_{(9)}\quad\nonumber ,\\
\d A^{\a\b\g}_{(10)} &=& 10 \de \L^{(\a\b\g)}_{(9)} -\tfrac{2}{3}
F^{(\a\b}_{(9)} \L^{\g)}_{(1)}
  + 32 F^{(\a}_{(3)} \L_{(7)}^{\b\g)} \label{IIBgt} \quad .
\label{bosgauge1}
\end{eqnarray}

The field-strengths, which are invariant
under the bosonic gauge transformations, are given by:
\begin{eqnarray}
F^\a_{(3)} &=& 3 \de A^\a_{(2)} \quad\nonumber ,\\
F_{(5)} &=& 5 \de A_{(4)} + \tfrac{5i}{8} \e_{\a\b} A^\a_{(2)} F^{\b}_{(3)}
\quad\nonumber ,\\
F^{\a}_{(7)} &=& 7 \de A^\a_{(6)} + 28 A^\a_{(2)} F_{(5)}
  - \tfrac{280}{3} F^\a_{(3)}A_{(4)} \quad\nonumber ,\\
F^{\a\b}_{(9)} &=& 9 \de A^{\a\b}_{(8)} + \tfrac{9}{4} F^{(\a}_{(7)}
A^{\b)}_{(2)}
 - \tfrac{63}{4} F^{(\a}_{(3)}A^{\b)}_{(6)} \quad\nonumber ,\\
F^{\a}_{(11)} &=& 11 \partial A^{\a}_{(10)} = 0\quad\nonumber ,\\
F^{\a\b\g}_{(11)} &=& 11 ( \partial A^{\a\b\g}_{(10)}
 -\tfrac{1}{3} F^{(\a \b}_{(9)} A^{\g)}_{(2)} + 4  F^{(\a}_{(3)}
A^{\b \g )}_{(8)}) = 0 \quad \label{lastcurv}.
\end{eqnarray}

It is clear that the bosonic transformations commute, because the
transformations $\delta A_{(2n)}$ contain only parameters and gauge
invariant curvatures. In other words, we have nonlinear transformation
rules and an Abelian gauge algebra.
Following \cite{Juliaetal} we write out the
curvatures in \eqref{abelian},
using \eqref{lastcurv}. Next, we
redefine the parameters $\L$ and $\S$ of the gauge transformations
such that the transformations only depend on $d\Lambda, d\S$, and not on
$\Lambda,\S$. After that, we redefine the bosonic gauge fields to make
the bosonic gauge transformations linear in the gauge fields.
Finally, we suitably rescale the fields and parameters to simplify the
form of the transformations. This leads to the following form for the
gauge transformations:
\bea \d A^\a_{(2)} &=& \Lambda^\a_{(2)}\quad\nonumber ,
\\
\d A_{(4)} &=& \Lambda_{(4)}+i\e_{\g\d}\L^\g_{(2)}A^\d_{(2)} \quad\nonumber ,
\\
\d A^\a_{(6)} &=& \L^\a_{(6)}
     +\L_{(4)}A^\a_{(2)} +\g \L^\a_{(2)} A_{(4)}
\quad\nonumber ,
\\
\d A^{\a\b}_{(8)} &=& \L^{\a\b}_{(8)}
  +\L^{(\a}_{(6)}\,A^{\b)}_{(2)} + \e \L^{(\a}_{(2)}A^{\b)}_{(6)} \quad\nonumber ,\\
\d A^{\a\b\g}_{(10)} &=& \L^{\a\b\g}_{(10)} +\L^{(\a \b}_{(8)}\,A^{\g)}_{(2)}
+ \m \L^{(\a}_{(2)}A^{\b\g)}_{(8)}\quad\nonumber ,\\
\d A^{\a}_{(10)} &=& \L^{\a}_{(10)}
\label{bosgauge2}
\eea
with
\be \label{values}
    \g=-2\,,\quad \e=-3\,,\quad \m=-4\ .
\ee
Note that, even though we have rescaled both fields and gauge parameters
we use the same notation as for the original fields in (\ref{IIBgt}), to avoid an
excess of complicated notation.
Also, in this subsection we use the notation
$\L_{(2n)}\equiv \partial\L_{(2n-1)}$, following \cite{Juliaetal}. The
three coefficients $\g,\e,\m$ can either be derived directly from
the supersymmetry algebra as explained, or be obtained by closure of
the bosonic gauge algebra. In either case we find the values given in
\eqref{values}.

So the structure we find is very rigid and, with our
requirements (parameters appear only with derivatives, linearity in
fields, non-trivial transformations), unique. The bosonic gauge
algebra in this form is given by the following commutation
relations: \bea \label{IIBalgebra}
   [ \d_{\tilde\L_{(2)}},\delta_{\L_{(2)}}] &=&
   \delta_{\L_{(4)}}\big(
   \L_{(4)}=-2\e_{\g\d}\tilde\L^\g_{(2)}\L^\d_{(2)}\big)\quad\,,
\nonumber\\
    {[} \d_{\tilde\L_{(2n)}},\delta_{\L_{(2)}}] &=&
    \delta_{\L_{(2n+2)}}\big(
    \L_{(2n+2)}=-(n+1)\tilde\L_{(2n)}\L_{(2)}\big)\ {\rm for}\ n>1
\quad\,. \eea In this formula we have suppressed the $SU(1,1)$ indices.

Note that, again, we use the same notation as for the original fields in
\ref{IIBgt}, to avoid an excess of complicated notation.
Thus the bosonic gauge algebra is rather special. In a sense, the
starting point is also rather special, because it is commutative. In
\cite{Juliaetal} it is suggested that nonabelian algebras, such as the
one we
obtained above are always related to commutative algebras,
as in our starting
point.\\
The above results suggest that it might  be possible to find a
basis for the fields of IIB supergravity, in which the supersymmetry
transformations of the gauge fields are linear in the gauge fields. This
is indeed the case. In fact, the supersymmetry transformations, as
presented in \cite{BdRKR}, already are in this form. To show this it
is convenient to denote the terms in  the
supersymmetry transformations of the bosonic gauge fields that
explicitly contains the gravitino or dilatino with   $\d_{F}$.
Using this notation we can write
out the supersymmetry rules  given in formulae (5.1) to (5.11)
in \cite{BdRKR} as:
 \bea
   \d A^\a_{(2)} &=& \d_F  A^\a_{(2)}\quad,
\nonumber\\
   \d A_{(4)}    &=& \d_F  A_{(4)} -
      \tfrac{3i}{8}\e_{\g\d}  A^\g_{(2)} \d_F A^\d_{(2)}  \quad,
\nonumber\\
   \d A^\a_{(6)} &=& \d_F A^\a_{(6)} +40 A_{(4)}\d_F A^\a_{(2)}
      -20 \d_F A_{(4)}  A^\a_{(2)} \quad,
\nonumber\\
   \d A^{\a\b}_{(8)} &=&  \d_F A^{\a\b}_{(8)}
    + \tfrac{21}{4} A^{(\a}_{(6)} \d_F A^{\b)}_{(2)}
    - \tfrac{7}{4} A^{(\a}_{(2)}\d_F A^{\b)}_{(6)}
   \quad,
\nonumber\\
   \d A^{\a\b\g}_{(10)} &=& \d_F A^{\a\b\g}_{(10)}
    -12  A^{(\a\b}_{(8)}\d_F A^{\g)}_{(2)}
    +3 A^{(\a}_{(2)} \d_F A^{\b\g)}_{(8)} \quad ,
\label{linearsusy}
\eea
so, there are no terms nonlinear in the gauge fields.

Note that the relative coefficients in (\ref{linearsusy}), i.e., between
$ A_{(4)}\d_F A^\a_{(2)}$ and $ \d_F A_{(4)}  A^\a_{(2)}$ etc.,
are $-2,\ -3$ and $-4$. The same coefficients occur in the
corresponding curvatures (\ref{lastcurv}), if the $F_{(n)}$
on the right-hand side are replaced by $n\partial A_{(n-1)}$.

The absence of terms of higher order in the gauge fields
in (\ref{linearsusy}), and the numerical correspondence with
(\ref{lastcurv}), can be understood from the requirement that
(\ref{lastcurv}) can be extended to a set of supercovariant curvatures.
The appearance of the same coefficients in (\ref{bosgauge2}) is
not surprising considering the close correspondence between
(\ref{bosgauge2}) and (\ref{lastcurv}).

\subsection{The IIA Algebra}

Our starting point for the IIA algebra are the bosonic gauge
transformations\footnote{Note that in the IIA case we work in
string frame and use a mostly plus metric.}
and field-strengths given in section 2. As in the IIB case, we write out the
curvatures in the variations of the potentials explicitly, and
redefine the parameters $\L$ and $\S$ of the gauge transformations
such that the transformations depend on $\partial\Lambda$ and $\partial
\S$\,
\footnote{We treat the mass parameter $G^{(0)}$ like a derivative for this
purpose, so $G^{(0)}\Lambda$ would, for example, also be of the
desired form. However, this makes the notation $\L_{(2n)}\equiv \partial
\L_{(2n-1)}$
which we used in the
previous subsection unpractical.},
but not on $\L$ and $\S$. The second step is to
redefine the bosonic gauge fields to make the bosonic gauge
transformations linear in the gauge fields. The last step is to
suitably rescale fields and parameters to simplify the form of the
transformations. We find: \bea
   \d B_2 &=& \partial \S_{(1)} \quad ,
\nonumber\\
   \d C_{(1)} &=& \partial\L_{(0)} - G_{(0)}\S_{(1)}\quad,
\nonumber\\
   \d C_{(3)} &=&  \partial\L_{(2)} - C_{(1)} \partial\S_{(1)}\quad,
\nonumber\\
   \d C_{(5)} &=&  \partial\L_{(4)} -  C_{(3)} \partial\S_{(1)}\quad,
\nonumber\\
   \d C_{(7)} &=&  \partial\L_{(6)} -  C_{(5)} \partial\S_{(1)}\quad,
\nonumber\\
   \d C_{(9)} &=&  \partial\L_{(8)} -  C_{(7)} \partial \S_{(1)}\quad,
\nonumber\\
   \d B_{(6)} &=&  \partial \S_{(5)} + G_{(0)} \L_{(6)}
  +\tfrac{1}{2}\big(-C_{(1)} \partial \L_{(4)} +  C_{(3)} \partial
 \L_{(2)}-C_{(5)} \partial \L_{(0)}
      +G_{(0)} C_{(5)} \S_{(1)}\big) \quad,
\nonumber\\
   \d B_{(8)} &=& \partial \S_{(7)} - G_{(0)}\L_{(8)}
 + \tfrac{2}{5}  B_{(6)} \partial \S_{(1)}
\nonumber\\
&& \quad +\tfrac{1}{5}\big(2 C_{(1)} \partial \L_{(6)} - C_{(3)} \partial
\L_{(4)}
   + C_{(7)} \partial \L_{(0)} - G_{(0)}  C_{(7)} \S_{(1)}\big)\quad ,
\nonumber\\
   \d D_{(10)} &=& \partial\S_{(9)} + G_{(0)} \L_{(10)} +\tfrac{5}{8} B_{(8)}
\partial \S_{(1)} +\tfrac{1}{16}\big(-5 C_{(1)}\partial\L_{(8)} + C_{(3)}
\partial\L_{(6)}
\nonumber\\
&&
 +C_{(5)}\partial\L_{(4)} -C_{(7)} \partial\L_{(2)}- C_{(9)}\partial\L_{(0)}
    + G_{(0)} C_{(9)} \S_{(1)} \big)\quad  \nonumber.
\eea
Note that, even though for simplicity our notation does not indicate it, the
fields and gauge parameters have been redefined and are not the same as those
of section \ref{IIA}.
This leads to the following algebra. On the RR forms we find only:
\bea
&&   [ \d_{\tilde\S_{(1)}}, \d_{\S_{(1)}} ] =
   \d_{\L_{(2)}}(\L_{(2)} = -G_{(0)}\tilde\S_{(1)}\S_{(1)})\quad ,
\label{S1S1}\\
&&   {[} \d_{\L_{(2k)}},\d_{\S_{(1)}} ] =
      \d_{\L_{(2k+2)}}\big(\L_{(2k+2)}= -\L_{(2k)}\partial\S_{(1)}\big)\quad ,
\label{L2kS1}\\
&&
     {[} \d_{\L_{(2k)}},\d_{\L_{(2l)}} ] = 0\quad .
\label{L2kL2l} \eea This algebra is extended once we
consider the action on the NSNS forms. For example, the commutator
(\ref{S1S1}) must also be realized on the NSNS fields. This is
indeed the case. The commutator (\ref{L2kS1}) is extended with a
$\d_{\S_{(2k+1)}}$ transformation: \bea
 {[} \d_{\L_{(2k)}},\d_{\S_{(1)}} ] &=&
  \d_{\L_{(2k+2)}}\big(\L_{(2k+2)} = -\L_{(2k)}\partial\S_{(1)}\big)
\nonumber\\
&&\quad
  +\d_{\S_{(2k+1)}}\big(\bar\S_{(2k+1)} = x_{2k}\, G_{(0)}\L_{(2k)}\S_{(1)}
\big)\quad ,
\eea with $x_6=1,\ x_8=-3/5$, and $x_{10}=6/16$. We also have: \bea
  {[}\d_{\S_{(2k+1)}},\d_{\S_{(1)}} ] &=&
  \d_{\S_{(2k+3)}}\big(\bar\S_{(2k+3)}=y_{2k+1}\,\S_{(2k+1)}\partial\S_{(1)}
\big)
\quad ,
\eea with $y_5=2/5,\ y_7=5/8$. Finally, many of the commutators
between two $\L$ transformations become nonzero and give a
$\S_{(2k+1)}$ transformation for $k>1$. We write these as \bea
   {[} \d_{\L_{(2k)}},\d_{\L_{(2l)}} ] &=& \d_{\S_{(2k+2l+1)}}
  \big(\bar\S_{(2k+2l+1)} = z_{2k,2l} \,(\L_{(2k)} \partial\L_{(2l)}
   - \L_{(2l)} \partial\L_{(2k)} \big)\quad,
\eea with $z_{4,0}=-1/2,\ z_{6,0}=3/10,\ z_{8,0}=-3/16,
z_{4,2}=1/10, z_{2,2}=1/4,\ z_{4,4}=1/32$. Other combinations
vanish.

Having established the form of the IIA and IIB bosonic gauge algebras
we are now in a position to discuss
an intriguing relation between these algebras and
the Kac-Moody algebra $E_{8}^{+++}$, which is
also called $E_{11}$. We first consider the IIB algebra, see
\eqref{IIBalgebra}.
Using an obvious notation this algebra has the following schematic form:

\be
{\rm IIB}\,:\hskip .5truecm
\boxed{[{\bf{2}},{\bf{2}}] = 4\,,\hskip
1truecm [{\bf{2}},4]=6\,,\hskip 1truecm [{\bf{2}},6]=8\,, \hskip .5truecm
\cdots}
\ee

We thus see that the gauge transformation $\Lambda_{(2)}^\alpha$
of the 2-form, indicated by $\bf{2}$ above, acts like a raising
operator in the sense that all the $2n$-form gauge transformations
$\Lambda_{(2n)}$ with $n>1$ can be obtained as multiple commutators of the
$\bf{2}$ transformation. This is reminiscent to a  similar structure
that occurs in $E_{11}$ in a rather different context.
For instance, in \cite{kleinschmidtwestschnakenburg} the
algebra $E_{11}$ was decomposed  in a particular way
with respect to $SL(10)$,
which should be thought of as the spacetime symmetry group.
This leads to the Dynkin diagram in figure 1.

\begin{center}\epsfig{file=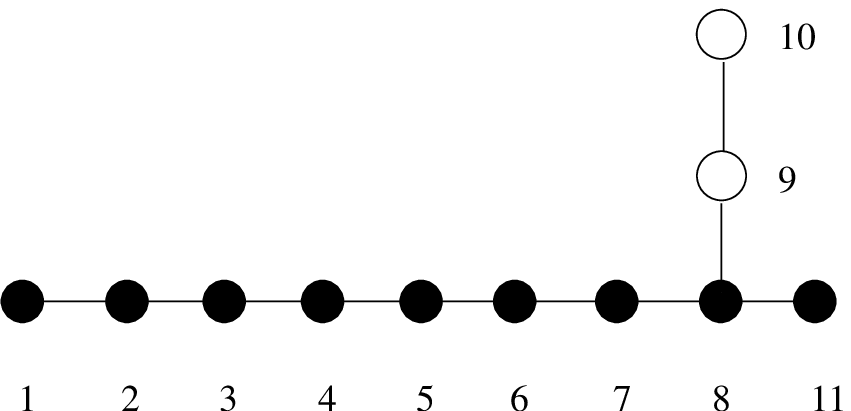,
width=0.4\textwidth}
\end{center}
\hskip 2truecm  Fig 1:\ \ The IIB decomposition of the $E_{11}$ Dynkin diagram.

\vskip .5truecm

The nine black dots represent the $SL(10)$ sub-algebra on which the
gravity sector is embedded.
It can be shown that the "lowest" irreducible representations arising for
this decomposition coincide with the fields
of IIB supergravity including the 10-form potentials, see
table 3 in appendix A.1  of~\cite{kleinschmidtwestschnakenburg}.
The way this works is that the two white dots act as two
raising operators
and the number of times they act corresponds to the ``level''
of the representation. In this way all representations can be obtained.
In our supergravity approach a similar thing happens in the bosonic
IIB gauge algebra where the
two white dots should be identified with the $\L_{(2)}^\a$
transformations. The fact that $\L_{(2)}^\a$ is a 2-form
follows in the Dynkin diagram from the presence of the two black dots
8 and 11. The fact that the 2-form gauge transformations transform as a
doublet under $SL(2,\mathbb{R})$ follows from
the presence of the two white dots 9 and 10.
The analogy is that in the same way as all relevant $SL(10)$
representations can be obtained by a multiple action of the
two raising operators all bosonic gauge transformations can be obtained
as a multiple commutator of the basic $\L_{(2)}^\a$ transformation.

We next consider the IIA algebra, see subsection 3.2,
 where a similar thing happens. Schematically the IIA bosonic gauge algebra
is given by

\be
{\rm IIA}\,:\hskip .5truecm
\boxed{[{\bf{1}},{\bf{1}}]=0\,, [{\bf{1}},{\bf{2}}]=3\,,
[{\bf{1}},3]=0\,,[{\bf{2}},3]=5\,,[{\bf{1}},5]=6\,,\hskip .5truecm \cdots}
\ee

We thus see that in this case the gauge transformations $\L_{(0)}$ and
$\S_{(1)}$, indicated by $\bf{1}$ and $\bf{2}$ above,
 act as two raising operators in the sense that all other gauge
transformations can be obtained as multiple commutators of $\bf{1}$ and
$\bf{2}$. This
corresponds to the level structure in
another decomposition of the $E_{11}$ Dynkin
diagram
with respect to $SL(10)$ (see Fig. 2) \cite{kleinschmidtwestschnakenburg}.

\begin{center}\epsfig{file=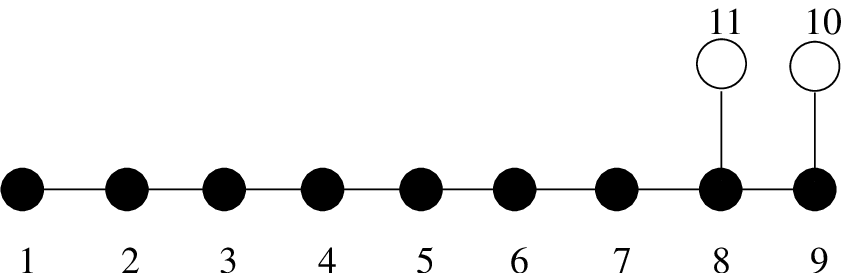,
width=0.4\textwidth}
\end{center}
\hskip 2truecm  Fig 2:\ \ The IIA decomposition of the $E_{11}$ Dynkin diagram.

\vskip .5truecm

Like in the IIB case the two white dots indicate the two
raising operators. However, in this case, they correspond to
a 2-form $\S_{(2)}$ (the 11 white dot) and a 1-form $\L_{(1)}$
(the 10 white dot). The calculation of the $SL(10)$ representations
leading to the fields of IIA supergravity including the 10-form
potentials can be found in table 2 and the corresponding table in
Appendix A.1 of
\cite{kleinschmidtwestschnakenburg}.
An explicit construction of the gauge algebra from the $E_{11}$
point of view has been given in
\cite{west10forms} and earlier works (see references in
\cite{west10forms}) .

Although the similarities between the IIA and IIB bosonic
gauge algebras and the predictions by $E_{11}$
are intriguing there are also striking differences. The
most important one is that the $E_{11}$ symmetries
predict many more $SL(10)$ representations whose interpretation from the
supergravity point of view are unclear at the moment.
Nevertheless we consider it remarkable that there is so much overlap
between the predictions of IIA/IIB supersymmetry and the bosonic
$E_{11}$ symmetry.

\section{M-theory\label{M-theory}}

It is natural to consider our results from an M-theory perspective.
It turns out that none of the two IIA 10-form potentials has a $d=11$
origin. It is well-known that the same is true for the RR 9-form potential.
This is related to the fact that massive IIA supergravity has no known
$d=11$ origin at the field theory level. We have independently verified that the
$d=11$ superalgebra does {\it not} allow the inclusion of an 11-form
potential.

It is interesting to see what happens with the bosonic gauge algebra of M-theory which
was investigated in \cite{Juliaetal}. The fields of the $d=11$
supergravity multiplet consist of a graviton $g_{\mu\nu}$, a 3-form
potential $C_{(3)}$ and a dual potential $C_{(6)}$. Using the
same notation as above the bosonic gauge algebra has
the following schematic form:

\be
\boxed{
[{\bf{3}},{\bf{3}}]={6}\,,\hskip 2truecm
[{\bf{3}},6]=0 \,.
}
\ee
\vskip .3truecm

In order to produce the $E_{11}$ structure we would like a rank 9
symmetry  to occur at the right-hand side of the $[{\bf{3}},6]$
commutator. However, there is no 9-form potential available in
$d=11$ supergravity. Instead, the $[\bf{3},6]$ commutator can also
give rise to a $(8,1)$-form which one could identify at the
linearized level with the $d=11$ dual graviton
\cite{Curtright:1980yk,west1,Hull:2001iu}. This is in fact predicted
by $E_{11}$ \cite{west1}\footnote{Actually, similar dual gravitons
are predicted to occur in the IIA and IIB case
\cite{kleinschmidtwestschnakenburg}.}. These  representations follow
from yet another decomposition of the $E_{11}$ Dynkin diagram in
terms of an $SL(11)$ bosonic subalgebra:
\begin{center}\epsfig{file=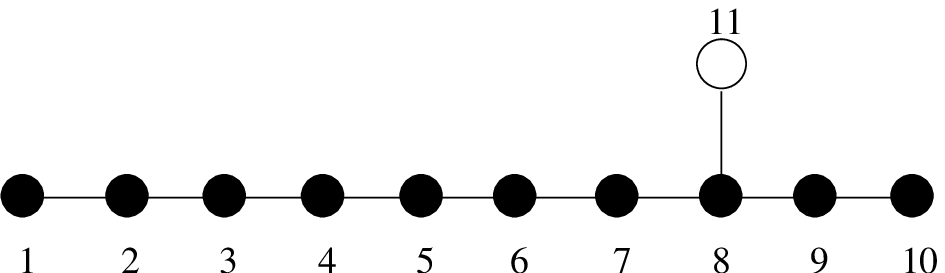,
width=0.4\textwidth}
\end{center}
\hskip 2truecm  Fig 3:\ \ The M-theory
decomposition of the $E_{11}$ Dynkin diagram.

\vskip .5truecm
The ten black dots correspond to the $SL(11)$ subalgebra and the
white dot indicates a single raising operator.
The fact that the gauge transformation
 is a 3-form follows from the three black dots
8, 9, and 10. The specific representations predicted by $E_{11}$
can be found in
table 1 of \cite{kleinschmidtwestschnakenburg}.

Extending dual gravitons to the nonlinear
level seems to be problematic \cite{Bekaert:2002uh}. It would
be interesting to reconsider this issue in the context of
(linearised) $d=11$ supergravity and the underlying $E_{11}$ structure.

\section{Conclusions\label{Concs}}

We have presented the supersymmetry and gauge transformations of a
completely democratic IIA supergravity theory. This has led to the
insight that IIA supergravity admits two distinct 10-form
potentials. In the massive version of the theory, which is naturally
included in our completely democratic formulation, one of the
10-forms, as well as the 6- and 8-forms can be gauged away. The
natural role of the 10-forms is to couple to 9-branes. We have shown
that the IIA theory may contain a kappa-symmetric 9-brane. The
consistency of such a 9-brane would require the presence of a
corresponding orientifold plane, along the lines of \cite{hull}.

The second part of this paper was concerned with the bosonic
gauge algebras, which are contained in the IIA and IIB theories. We
have presented a formulation in which the transformation rules are
linear in the gauge fields and the bosonic gauge algebras are Non-Abelian.
These algebras turn out to be the bosonic
algebras of \cite{Juliaetal}, extended with 10-forms. These algebras
 also play a role in the conjectured $E_{11}$ symmetry, which might underly
M-theory \cite{west1,west3}.

\section*{Acknowledgements}

We would like to thank Axel Kleinschmidt and Peter West for
useful discussions.
F.R. would like to thank the University of Groningen for
hospitality. E.B., S.K. and M. de R. are supported by the European
Commission FP6 program MRTN-CT-2004-005104 in which E.B., S.K. and
M. de R. are associated to Utrecht university. S.K. is supported by
a Postdoc-fellowship of the German Academic Exchange Service (DAAD).
F.R. is supported by a European Commission Marie Curie Postdoctoral
Fellowship, Contract MEIF-CT-2003-500308. F.R. also thanks
INFN for support, and the Department of Phsyics at University of Rome
``Tor Vergata'' for hospitality. The work of E.B. and T.O. is
partially supported by the Spanish grant BFM2003-01090.

\end{document}